%% file: main.tex
\documentclass[manuscript]{acmart} % for arxiv

\usepackage{graphics}
\usepackage{soul}
\usepackage{multirow}

\AtBeginDocument{%
  }

\copyrightyear{2025} 
\acmYear{2025} 
\setcopyright{acmlicensed}\acmConference[CHI '25]{CHI Conference on Human Factors in Computing Systems}{April 26-May 1, 2025}{Yokohama, Japan}
\acmBooktitle{CHI Conference on Human Factors in Computing Systems (CHI '25), April 26-May 1, 2025, Yokohama, Japan}
\acmDOI{10.1145/3706598.3713958}
\acmISBN{979-8-4007-1394-1/25/04}

\AtBeginDocument{\colorlet{defaultcolor}{.}}

\newif{\ifhidecomments}
\hidecommentstrue
\ifhidecomments
    \newcommand{\rev}[1]{\textcolor{defaultcolor}{#1}}
\else
    \newcommand{\rev}[1]{\textcolor{purple}{#1}}
\fi

\begin{document}

\title{Talking About the Assumption in the Room}

\author{Ramaravind Kommiya Mothilal}
\email{ram.mothilal@mail.utoronto.ca}
\affiliation{%
  \institution{University of Toronto}
  \country{Canada}
}

\author{Faisal M. Lalani}
\email{faisalmlalani@gmail.com}
\affiliation{%
  \institution{University of Illinois Urbana-Champaign}
  \country{USA}
}

\author{Syed Ishtiaque Ahmed}
\email{ishtiaque@cs.toronto.edu}
\authornotemark[1]
\affiliation{%
  \institution{University of Toronto}
  \country{Canada}
}

\author{Shion Guha}
\email{shion.guha@utoronto.ca}
\authornote{Served as co-supervisors and contributed equally to this work.}
\affiliation{%
  \institution{University of Toronto}
  \country{Canada}
}

\author{Sharifa Sultana}
\email{sharifas@illinois.edu}
\affiliation{%
  \institution{University of Illinois Urbana-Champaign}
  \country{USA}
}

\renewcommand{\shortauthors}{Kommiya Mothilal et al.}


% The presence and controversy of assumptions in technical ecosystems is well-established in responsible AI discourse. What is seldom touched upon is the conceptualization around assumptions, and how AI practitioners identify and handle them throughout their workflows. We present a theoretical framework of how these practitioners generally navigate assumptions and offer recommendations on how to incorporate reflective practices that allow organizations to better utilize these assumptions in avoiding potential harms.

\begin{abstract}
The reference to \textit{assumptions} in how practitioners use or interact with machine learning (ML) systems is ubiquitous in HCI and responsible ML discourse. However, what remains unclear from prior works is the conceptualization of assumptions and how practitioners identify and handle assumptions throughout their workflows. This leads to confusion about what assumptions are and what needs to be done with them. We use the concept of an \textit{argument} from Informal Logic, a branch of Philosophy, to offer a new perspective to understand and explicate the confusions surrounding assumptions. Through semi-structured interviews with 22 ML practitioners, we find what contributes most to these confusions is how \textit{independently} assumptions are constructed, how \textit{reactively} and \textit{reflectively} they are handled, and how \textit{nebulously} they are recorded. Our study brings the peripheral discussion of assumptions in ML to the center and presents recommendations for practitioners to better think about and work with assumptions. 
\end{abstract}





\maketitle

\input{1.introduction}

\input{2.literature}
\input{3.methods} 
\input{4.findings}
\input{5.discussion}
\bibliographystyle{ACM-Reference-Format}
\bibliography{references}

\appendix
% \clearpage
\input{6.appendix}

\end{document}
\endinput
%%
%% End of file `sample-sigconf-authordraft.tex'.

%% file: 1.introduction.tex
\section{Introduction}

Imagine a technical product manager, Meena, overseeing the development of a fictional healthcare application predicting the efficacy of a recently manufactured tuberculosis drug. Distribution data is collected from a trusted federal agency, demographics are cleaned and sorted through manual categorization, and trends are validated by organizational expectations. The model finds that the drug is most likely to reach and be effective for young adults in suburban neighborhoods. Meena, when writing out the limitations section of the model documentation, reflects on personal choices made within the process. She reveals that the federal agency was the only considered data source due to its colloquial familiarity, the data annotations for age ranges reflected the bias of a younger development team, and the results reinforced positive business forecasts. 

As a reader, what immediately stands out to you in Meena's disclosure? Perhaps you hone in on the subjectivity of what constitutes the social category of ``young adult.'' Or perhaps you deliberate on the association with government procedures or business interests as precarious infusions of hidden agendas. Perhaps, with alternative personal decision-making, the model may have also found that the drug showed significant side effects for older patients in rural settings with less government oversight.

What you considered in this hypothetical story were \textit{assumptions}. Their recognition, particularly in machine learning (ML) ecosystems, has become prevalent through the integration of trust and safety teams and increased adoption of responsible ML frameworks \rev{\cite{mitchell2021algorithmic,aragon2022human,srivastava2019mathematical}.} The presence of assumptions in any workflow is inevitable---they drive institutional motivations \rev{\cite{cheng2022child,holten2020shifting,stevenson2019algorithmic,saxena2024algorithmic}}, inform model specifications \rev{\cite{robertson2021modeling,michel2023recalibrating,obermeyer2019dissecting,kilbertusTraditionalAssumptionsFair2020}}, and allow the project to exist in the first place \cite{mitchell2021algorithmic,kilbertus2021beyond,malik2020hierarchy,saxena2021framework}. What is seldom acknowledged is their nuance and how workflow interactions often fail to capture that nuance through reflective interventions. This gap perpetuates a logic of distancing sociotechnical enablers from purely technical ones, leading practitioners to form a superficial understanding and adversarial response to them \rev{\cite{yildirim2023investigating,rakovaWhereResponsibleAI2021,kaurInterpretingInterpretabilityUnderstanding2020,mcconvey2024not}}. In essence, assumptions are employed one way or another, and their manifestation is a key part of any ML workflow. However, the centrality of their effect conflicts with the marginalization of their conceptualization.

% In this qualitative study, we explore this marginalization through the lens of argumentation theory in \textit{Critical Thinking} and \textit{Informal Logic}, two fields that define assumptions as the premise of \textit{arguments} an individual posits in the service of achieving a goal. 
In this qualitative study, we explore this marginalization through the lens of argumentation theory in the field of  Informal Logic, where assumptions are defined as the premises of an \textit{argument} an individual posits in the service of achieving a goal \rev{\cite{hitchcock2007informal,johnson2012manifest,ennis2006probably}.} 
These arguments culminate in a conclusion that postulates a pillar of reasoning for the workflow to continue to exist \rev{\cite{goddu2018against}}. But because the rationale is composed of a weak relationship between the individual making the argument and their \textit{understanding} of the assumptive nature of its premises, there is often confusion in propagating any result and alleviating concern about harms. \textbf{The primary question we therefore ask is what contributes to this confusion about assumptions in an ML context}. Through unpacking the factors that allow assumptions to persist in an organization unchecked, this research attempts to explain how ML practitioners first understand assumptions and second, deal with them. Ultimately, we answer our inquiry by offering dual characterizations of the former and breaking down the procedural elements that may circumvent or subvert the latter.

While past works have heavily implied that assumptions play a crucial role in ML, few have positioned them directly in relation to workflow requirements and constraints (section \ref{rel:periphery}). As we describe in section \ref{sec:ontology}, assumptions can be ontologically categorized into an \textit{independent construction}---existing as an individual entity that maneuvers alongside a workflow---or a \textit{relative construction}---lurking within the fabric of technical or business processes. Both present implications for how assumptions are identified by practitioners, if at all, and what constitutes appropriate reactions when they reveal internal or external consequences. Section \ref{sec:procedure} elaborates on organizational patterns that actualize these constructions: first, we investigate how the realities of ML in practice contradict the constituents necessary to examine assumptions reflectively; second, we show how in-place structures intended to deal with assumptions are inadequate. Finally, in section \ref{sec:discussion}, we culminate the themes established prior to outline how practitioners can better articulate assumptions and set up internal processes to work \textit{with} assumptions rather than \textit{in spite} of them. In summary, our work contributes the following:

\begin{itemize}
    \item An investigation of an overlooked and consequential phenomenon in HCI and responsible ML discourse---the factors that contribute to the \textit{confusion} around assumptions in machine learning.
    \item A deconstruction of the vague conceptualizations and inconsistent views around assumptions and how they present themselves in documentation around limitations or requirements.
    \item An overview of how an assumption traverses the phases of a typical ML workflow and leads to downstream effects in data pipelines and modeling while possibly begetting more uncertainties.
    \item A framework to remedy the articulation and informal recording of assumptions by prompting practitioners to reflect on their structure and logic.
\end{itemize}

%% file: 2.literature.tex
\section{Related Literature}

Making simplifying assumptions to operate at a required abstract level is central to machine learning practice \cite{selbst2019fairness,saitta2013abstraction}. Abstraction inherently involves making assumptions about what is necessary and what is not. For instance, ignoring certain features or choosing a specific representation in data abstracts out certain social contexts and interactions that may be assumed as nonessential. As much as these traits have contributed to the rise of ML applications in diverse domains, the last decade has seen an increasing number of concerns arising from abstraction and assumptions made about human behaviors and characteristics in automated decision-making systems \cite{benjamin2019race,noble2018algorithms,o2017weapons,eubanks2018automating}. While assumptions form an essential constituent of many prior works on how practitioners use ML, in section \ref{rel:periphery}, we review how \rev{prior works in HCI and related disciplines} often place assumptions on the periphery. Then, in section \ref{rel:core}, we discuss how the concept of an \textit{argument} in Informal Logic can offer a new perspective to think about and act on the confusions surrounding assumptions in ML.

\subsection{Assumptions on the Periphery}
\label{rel:periphery}

% \rev{\textbf{Assumptions as Marginal Disruptor.}} 
Prior works in HCI, ML fairness, and AI ethics have extensively looked into how practitioners use ML systems and interact with different phases of an ML workflow (for e.g., \rev{\cite{zhangHowDataScience2020,yang2018investigating,wang2023designing,muller2020interrogating}}.) Many of these works have brief discussions about assumptions or at least mention the word ``assumption'' when analyzing practitioners' interactions with ML-based systems. However, most of these references assign a marginal causal agency to assumptions for disrupting a desired state or chain of actions: some common references to assumptions include phrases such as \textit{``The result is often erroneous assumptions [made by practitioners] about what users would want from AI.''} \rev{\cite[p.~12]{subramonyam2022human}}, \textit{``...they assumed that succinct answers were sufficient.''} (indicating undesired documentation practices) \rev{\cite[p.~17]{heger2022understanding}}, and \textit{``participants who assumed sex was a sensitive feature attempted to mitigate biases in the ML pipeline by simply removing...''} (explaining undesired actions sequence) \rev{\cite[p.~5]{dengExploringHowMachine2022}}.

\rev{In addition to model efficiency,} the desired state or actions discussed in these prior works often revolve around sociotechnical concerns related to fairness, transparency, or collaboration. For instance, some prior works refer to practitioners' assumptions as one of the factors hampering their collaborative efforts with stakeholders of different technical backgrounds \cite{yang2018investigating,varanasi2023currently,wang2023designing}; a few others discuss how assumptions distort practitioners' understanding of ethical or fairness issues \cite{boyd2021datasheets,dengExploringHowMachine2022,aragon2022human,jarrahi2022principles}. 
\rev{In all of these works, assumptions are often discussed peripherally to the main discussion about a desired state or action, such as efficiency or fairness. 
Specifically, along with other factors such as institutional constraints and incentives, assumptions are discussed as one of the factors that affect practitioners' attainment of a goal.}

% Though assumptions are only superficially discussed per se in studying practitioners' usage, 
% \smallskip
% \noindent \textbf{Assumptions as Object of Enumeration.}
\rev{As assumptions are often discussed in terms of their effect,}
their influence in practice is well-appreciated in responsible ML discourse \cite{mitchell2021algorithmic,jarrahi2022principles,aragon2022human,malik2020hierarchy}. 
Consequently, research in HCI and responsible ML has developed numerous toolkits \cite{wong2023seeing} (refer to frameworks, guidelines, etc.) to invoke assumptions in everyday practice \rev{\cite{Sadek2024,Lavin2022,Smith_undated,rismani2023plane}}. There are also a significant number of these toolkits---some very popular such as model cards \cite{mitchellModelCardsModel2019} or datasheets \cite{gebruDatasheetsDatasets2021}---that do not include direct interrogative questions to invoke assumptions, but rather frame questions on various phases of an ML workflow intended to unearth hidden assumptions \rev{\cite{Pushkarna2022,Kawakami2024,Raji2020,elsayed2023responsible}}. 
\rev{In a recent work, \citet{kommiya2024towards} recommend that practitioners deliberately connect, in addition to listing down, the known assumptions to the desired states.}   
However, most of \rev{these} toolkits \rev{and guidelines} are only suggestive, simply prompting the practitioners to list down and document assumptions in some form. While the authors of these toolkits argue that this process intrinsically will help mitigate the undesired consequences of making assumptions, it remains unclear how practitioners actually \textit{conceptualize} and \textit{work through assumptions} when prompted with ``what are the assumptions'' or ``list down the assumptions'' type questions. Though one of the primary intentions of these toolkits is to make practitioners elicit and reflect on the assumptions, the methodology adopted in the toolkits gives meager attention to \rev{the practical steps that practitioners undertake during} assumption articulation, identification, and handling. 

% -- also consequently a substantive number of works disuxss waht assumptions are missed and ignored by practitioner 
% -- consequently enumeration and recording 

Overall, prior works in HCI and responsible ML mostly make peripheral references to assumptions \rev{as objects of enumeration to avoid disrupting a desired state or action.
Several studies provide a review of common assumptions that practitioners ignore or miss during statistical modeling \cite{wang2022against,mitchell2021algorithmic,malik2020hierarchy}.
However, it remains unclear if assumptions in ML can be objectively and uniformly viewed and acted upon, especially when practitioners work with diverse stakeholders to build an ML system.
Hence, to supplement prior works, we examine this fundamental problem of confusion and uncertainties that practitioners might face when conceptualizing and working through assumptions.}
% Overall, prior works in HCI and ML mostly make peripheral references to assumptions in studying practitioners' use of ML systems and in the development of toolkits. 
To our knowledge, there is no work in an ML context that places assumptions in the center stage and investigates the surrounding confusions. In the next section, we introduce \rev{the concept of an \textit{argument}} from philosophical thinking to the HCI community to (re)think about assumptions in ML.

% Finally, we also find several prior works that make no or sparse reference to assumptions  
% With recent focus on improving the regulation on AI systems, governance and risk management framework such as NIST's RMF, 

% - overally it seems like assumptions are on the periphery and not getging enough attneiont

% - how in these tools - while some encourage reflectivess - but do not go deeper,
% - others just do not enocurage reflective practice or gives rise to further assumptions and causes confusion
    
% c. So this sidelining and unclarity in what do people do with assumptions encouraged us to ask our RQ.
% Our findings show that this practitoiners face confusion and unclarity concerning assumptions due to this current setup and infrastrucutre they are in and have

% - last para - many works do not mention assmtipns at all or use other terms but implitly they mean assumptions - so the result is confusion
%     - data workflow is a good example
    
% Prior work has discussed how ML and data practitioners draw on their beliefs and experiences to make subjective decisions at various stages of an ML lifecycle (cite).
% Our findings contribute to and reinterpret this line of work through the assumptions and the surrounding uncertainties and tensions that (a) prompts practitioners to make those subjective decisions, and (b) practitioners assess in others' interpretations and actions.

\subsection{Bringing Assumptions to the Center}
\label{rel:core}
The sub-field of \textit{Informal Logic} in Philosophy emerged as a response to the inefficiency of tools and criteria of formal \textit{Logic}\footnote{The prefix ``informal'' in Informal Logic is contentious and it is sometimes argued as a formal enterprise in a different sense \cite{woods2000philosophical,levi2000defense,johnson1999relation}.} in analyzing and evaluating natural language discursive arguments \cite{blair2012informal,walton2008informal,scriven1980philosophical,scriven1976reasoning}.
Informal Logic appreciates the structural complexity of everyday language use, the formulation of unstated assumptions, and the epistemological questions surrounding argumentation, among others, that analytic and normative tools of Logic ignores or over-simplifies, distorting the meaning of arguments \cite{anthony2015informal,johnson2014rise}. 
Arguments are also extensively studied in \textit{Critical Thinking}, often associated with Informal Logic \cite{weinstein1990towards,johnson2012informal,crews2007critical}, that studies a mode of thinking about an object involving active interpretation, clear articulation and analysis of reasons, assumptions and conclusions, logical evaluation of explanations and evidences, self-regulation, and holding a disposition to use the above-mentioned skills \rev{\cite{fisher1997critical,hitchcock2018critical,emis1962concept,facione1990critical}}.
% While critical thinking can also be about information, communication, and observations, a large overlap between Critical Thinking and Informal Logic lies in their focus on formulating, analyzing, and evaluating \textit{arguments} presented in some form (written, spoken, pictorial, etc. or multi-modal \cite{groarke2015going}.)

In this study, we refer to the conception of an \textit{argument} put forward by \citet{hitchcock2021concept} and review prior works to establish a connection between an assumptions and an argument \cite{kingsbury2002teaching,hitchcock2007informal,govier1992good,goddu2009refining}. By referring to the structuring of arguments as discussed in Informal Logic, we situate assumptions as core elements of arguments that ML practitioners make or engage in \textit{implicitly} or \textit{explicitly}.
We use this assumption-argument paradigm to offer a new perspective to understand and explicate the confusions surrounding assumptions in ML for two reasons. First, assumptions do not exist in a vacuum; they exist as part of arguments that are expressed or implied \rev{\cite{plumer2017presumptions,delin1994assumption,ennis1982identifying}}.
Therefore, critically analyzing the structure of an argument can explain \textit{how} and \textit{why} assumptions are made and what contributes to the confusion. Second, when analyzing an assumption and the surrounding confusion, we are essentially making arguments ourselves to critically think how assumptions shape a resulting argument \cite{berman2001opening,brookfield1992uncovering,delin1994assumption,ennis1982identifying}.

\rev{\citet[p.~105]{hitchcock2021concept}} formulates a simple argument as a premise-conclusion complex as follows:
\begin{quote}
    \textit{A simple argument consists of one or more of the types of expression that can function as reasons, a ``target'' (any type of expression), and an indicator of whether the reasons count for or against the target.}
\end{quote}
Reasons in the above definition refer to the \textit{premises} of an argument that perform a specific function: they commit the author of an argument to \textit{``something's being the case''} either assertively or hypothetically \rev{\cite[p.~10]{hitchcock2007informal,searle1976classification}}. 
% \cite{hitchcock2007informal,searle1976classification}.
In other words, premises constitute the propositions and the accompanying intention (or the illocutionary act \cite{searle1975taxonomy}, in philosophical terminology). 
For instance, when we use ``suppose the data is not representative'' as a premise for an argument, we express the proposition that the data is not representative \textit{hypothetically}.  The \textit{target} or conclusion is also a proposition but can be an illocutionary act type of different kinds, including a directive, a commissive, an expressive, and a declarative \cite{toulmin1958uses,ennis2006probably}. Finally, arguments can also be complex where the premise of one argument can be the target of another and so on \cite{hitchcock2021concept}.

When an ML practitioner makes or uses an assumption, it is often made or used \textit{for} a particular purpose.
For instance, when a feature is assumed to be unnecessary, it is often for realizing a particular objective such as to reduce the complexity of feature space. 
Similarly, when a performance metric is chosen, it is done so that optimized model outcomes are relevant for decision-subjects. The structure of an argument, as described above, then suggests that assumptions can be perceived as premises for attaining a target.
\citet{ennis1982identifying} discuss how these premise-type assumptions back-up or fill gaps for realizing the conclusion, and so the falsity of these premise-type assumptions weakens the support provided for the target. In other words, assumptions now become an essential component of an argument that a practitioner makes or implies. We recognize that there could be other forms of arguments, such as using vivid descriptions to display an identity or marching in protests \cite{jacobs2000rhetoric,hample2015arguing}, but we interpret the actions and expressions of ML practitioners as a premise-conclusion complex in this study and leave further exploration to future studies.
We also recognize the possibilities of interpreting different assumptions in an ML workflow as categories other than premises, such as conclusions and presuppositions, which might require a new lens to analyze \cite{walton2008argumentation,ennis1982identifying}.

Now, there can be situations where a practitioner may \textit{only} be making a premise-type assumption, but an analyst will be the one who is inferring the argument and making a distinction between premise and target. 
For instance, an ML developer can exclude certain text sources from the data and proceed with training their language model, but it is the safety expert in their organization who actually attempts to dissect the reasoning behind the data exclusion assumption\footnote{\rev{Prior works in HCI and responsible ML often do not make a distinction between first- and higher-order assumptions. In other words, assumptions made by the practitioners are not distinguished from those that are interpreted by the authors or someone else. Our point is not to doubt the inferential validity of these works but instead call attention to the complexity of assumptions, which may influence how they are examined and handled \cite{berman2001opening,atkin2017investigating,korzybski1958science}.}}. 
In other situations, a practitioner might need an assumption that they did not explicitly use, but which an analyst could infer. Or practitioners may not need an assumption but could have unintentionally used an assumption in making an argument. It is important to note that arguments are not necessarily localized to what practitioners do or write about in their documentations and reports, instead any (un)intentional action performed by a practitioner, such as choosing a specific algorithm, can be interpreted as an argument whose assumptions could be unearthed by an analyst \cite{kjeldsen2015study,groarke2015going}. Overall, this \textit{premise-target lens of an argument} deconstructs an assumption to understand why it was made (by identifying and relating it to the target) and how it was made (for instance, implicitly or explicitly), and thereby can support us in understanding how and why confusions exist around assumptions in ML practice.

% \textbf{relevance to hci}

% - what is assumtion - thinking
% - analyzing an assumption angle - critical and reflective thinking bit in papers

% - argument not only made but also had or engaged - this in itself has arguments made - implied argument - doc can be seen as that

% - now, conclusions can also be assumptions - but usually straightforward and leae to future work
% - explain why analyzing arguments help us uncover confusions about assumptions

% - taxomony of presmie tyle assumptions
% - how tools of critical thinking helps us as a method and as a theory

% In philosophical language, a premise is a illocutionary act of a proposition, either expressed assertively or hypothetically. 
% Consider two simple arguments, A1 and A2 below, inspired from prior philosophical works on this topic:
% \begin{itemize}
%     \item \textit{A1: Suppose that the data is not representative of the target population. Then it makes sense to add more data points.} 
%     \item \textit{A2: The data is not representative of the target population. So it makes sense to add more data points.}
% \end{itemize}
% Though the premises in both A1 and A2 different, they express the same proposition that the data is not representative.
% The difference lies in the illocutionary act performed by the 

%% file: 3.methods.tex
\section{Methodology}

\smallskip
\noindent \textbf{Participant }\rev{\textbf{Recruitment and Demographics.}} Once receiving approval from our institutions' ethics boards, we posted an open call for participants in several AI-oriented online communities \rev{on Slack and LinkedIn}. The call invited practitioners involved in some capacity with the research, development, design, or implementation of ML to participate in in-depth qualitative interviews on how they conceptualize, identify, and handle assumptions within their work. 52 individuals responded to our call, out of which we recruited 22 respondents for remote semi-structured interviews through purposive sampling \cite{sharma2017pros}. While this may not yield a statistically representative sample, it still allowed us to explore rich and unique insights into the experiences of the participants we felt most capable of answering the research questions in our study \cite{sharma2017pros,roy2015sampling}. Those who demonstrated significant experience working on ML projects, either as developers, data scientists, or product managers, as well as individuals closely involved with responsible ML artifacts, were ultimately chosen to participate in our interviews.
\rev{Most of our participants were from Global North locations and identified as males. Table \ref{tab:demographics} provides more details about our participants.}

\begin{table*}[]
\centering
\begin{tabular}{|l|l|}
\hline
\textbf{Dimension} & \textbf{Distribution} \\ \hline
Gender & Male: 16, Female: 6 \\ \hline
Region & Global North: 18, Global South: 4 \\ \hline
Role & ML/Data Engineer: 7, ML/Data Scientist: 6,  Management: 5, \\ 
 & Others (Designer/Ethicist/Academic): 4, \\ \hline
Organization Type & Tech company: 10, NGO/Civil Society: 5, Consulting: 4, \\
& Academia: 2, Government: 1 \\ \hline
\end{tabular}
\caption{\rev{Participant Demographics}}
\label{tab:demographics}
\end{table*}

\smallskip
\noindent \textbf{Interview }\rev{\textbf{Design.}} \citet{brookfield1992uncovering} emphasizes that the key to uncovering assumptions lies in analyzing the lived experience of the assumer in order to embed a specific practice within a realistic context. This motivated how we framed our questions to be reflective, allowing participants to answer in a way that stepped outside their typical frames of reference and assess their assumptions by explicitly thinking about them. The questions were also designed to explore participants' experiences without presuming outcomes while allowing participants to refute our own underlying assumptions \cite{kvale2009interviews}. The downside of this direct approach is that unconscious assumptions---the ones that inform a participant's intuition without them being privy to their existence and persistence---may fall through. 
To remedy this, we offered a second part of the interview in which we extracted specific phrases from model documentation of three popular large language models--- PaLM 2 \cite{anil2023palm}, BLOOM \cite{le2023bloom}, and Llama 2 \cite{touvron2023llama}---and asked participants to vocally analyze them. We chose these models as they varied across different dimensions of openness \cite{liesenfeld2024rethinking}. The sample texts were selected because they most directly offered an argument that follows a typical premise-conclusion structure with deliberate non-technical language that may prompt confusion at first glance. The samples are provided in appendix \ref{appendix}.

Our approach in framing the lifecycle of an assumption by inquiring about how it is conceptualized\footnote{\rev{Some readers may wonder why we focus only on conceptualization and not consider the \textit{operationalization} of an assumption. However, in our experience and based on our interviews with practitioners, ML stakeholders do not operationalize the construct ``assumption'' in practice but operationalize only the content of a specific assumption (e.g., the usage of ``representative'' in the assumption ``this data is representative''). In this view, assumptions function at a meta-level as discussed in prior works in Critical Thinking and Informal Logic (section \ref{rel:core}), and so we focus only on the conceptualization of assumptions in this work to uncover the confusion associated with the practical use of this term in ML. We leave alternative explorations to future work.}}, then identified, then handled aligns with \citet{berman2001opening}'s breakdown of an assumption as a single entity composed of assuming, feeling, thinking, and behaving. By organizing questions through assessing \textit{functions} of assumptions rather than conveying them holistically, we are able to easily distinguish what specific elements contribute to confusion around assumptions, and how participants react to that confusion. Furthermore, following the logic that initial assumptions are likely to predicate how future assumptions are handled, we attempted to frame questions in a way that allowed us to form a narrative of a participant's assumptions.

Questions are also informed by our personal experience in ML ecosystems, aligning with established practices in \textit{reflexive} qualitative research \cite{berger2015now}. The idea of assumptions being present in technical ecosystems and the motivation for the study in assessing their influence is driven by our own observations working within the space and examining it from a critical lens derived from our past and current positions as responsible ML researchers. We make this position explicit to enhance the rigor, credibility, and trustworthiness of the study and allow readers to understand the lens through which we interpreted responses.

\smallskip
\noindent \rev{\textbf{Interview Procedure.}}
\rev{The interview guide was developed by the first and second authors and was thoroughly discussed and approved by all authors. We share our complete interview guide in Appendix \ref{interview}.
We sent our consent letters ahead of the interviews and gave our participants the option of either returning the signed letter or providing verbal consent during the interview.
All our interviews were conducted in English via Zoom. While the first and second authors conducted 9 interviews together, the first author conducted 12 interviews independently, and the second author conducted 1 independently. 
We recorded our calls upon consent and manually took notes of participants who were uncomfortable with recording. 
Our participants were given the option to exit the interviews whenever they needed. 
Our interviews lasted for 60 minutes on average. We compensated participants with 30\$ for their time and contribution.}    

\smallskip
\noindent \rev{\textbf{Data }}\textbf{Analysis.} Our virtual interviews yielded approximately 25 hours of recorded audio, paired with auto-generated transcripts from Zoom. \rev{The interview data and notes were stored in the first author's institutional cloud storage.} 
\rev{As described in our interview design, our questions were broadly framed to extract how assumptions are perceived, identified, handled, and used in practice.}
Given the nature of our more open questions, we employed interpretative and descriptive qualitative analysis \cite{merriam2019qualitative} to decipher \rev{insights} within the transcribed responses. 
\rev{The first and second authors conducted the bulk of the data analysis, and the final themes were discussed and finalized among all authors. The analysis began with multiple readings of the transcripts followed by open-coding on the transcribed data, independently and manually, by the first two authors. They then iteratively went through each other's codes manually, extracted and recorded commonalities, cross-checked with one another for reliability, and finalized the codes after resolving critical disagreements by open discussion.}

\rev{In the next phase, the codes were interpreted through the assumption argument lens (section \ref{rel:core}), mapped, categorized, and structured into themes and sub-themes over multiple iterations.} 
\rev{For instance, several sub-themes such as ``forgotton assumptions'' and ``recording style'' were grouped into one of the main themes, ``informal documentation.'' These sub-themes were created by grouping several codes that revolved around how practitioners noted down their and others' assumptions. 
Further, while some sub-themes, such as ``chained assumptions'' and ``granularity'' had overlapping codes, we categorized these sub-themes into distinct themes (elaborated in sections \ref{subsec:integrate} and \ref{subsec:doc} respectively) as it offers a better frame to understand the confusions around assumptions.
Overall, as key takeaways were found around how participants personally and professionally interacted with assumptions, we were able to form ontological distinctions, procedural inconsistencies, and other confusing elements that helped us craft clear constructions of an assumption, how workflows perpetuate unchecked assumptions, and what practitioners (can) do about it.  
Our findings in section \ref{sec:findings} reflects how we inferred and organized the main themes in our data.
}
% - Answers were categorized in themes - how participants defined and identified assumptions, what givens they had going into the ML process, how they handled assumptions, and how they responded to the case study.

% \smallskip
\noindent \textbf{Limitations.} A study about assumptions will naturally possess a few assumptions itself. First, the premise of the study requires a consensus between the authors and the participants that assumptions in an ML workflow have a significance that needs to be addressed, potentially influencing answers toward having a more proactive stance toward them. Second, the samples chosen in the case study portion of the interviews were pointers extracted from lengthier and more contextualized model documentation; our selection was informed by our own assumptions about what may elicit rich responses. The samples shown were also the same for all participants. While this provided an equal frame of reference, future works could reinforce our findings through comparing similar perceptions in more diverse samples.

%% file: 4.findings.tex
\section{Findings}
\label{sec:findings}
We find that the confusions concerning assumptions in ML largely revolve around what assumptions are (section \ref{sec:ontology}) and what is being done about them (section \ref{sec:procedure}). While many confusions are due to vague conceptualizations of assumptions, institutional fragmentation, and a general lack of clarity on response, others stem from holding unique and inconsistent views and procedures that deal with assumptions. Further, practitioners differ in their characterization of assumptions based on the role they take: whether they are the ones assuming (\textit{assumer}) or analyzing (\textit{analyst}) something. \rev{Table 1 summarizes our findings.}

\begin{table*}[]
\centering
\resizebox{\textwidth}{!}{%
\begin{tabular}{|l||l|l|l|}
\hline
\textbf{Axis of Confusion} & \textbf{Theme} & \textbf{Key Takeaways} & \textbf{Premise-Target Lens} \\ \hline
\multirow{2}{*}{\begin{tabular}[c]{@{}l@{}} Conceptualization of \\ what assumptions are\end{tabular}} & \begin{tabular}[c]{@{}l@{}}Independent\\ construction\end{tabular} & \begin{tabular}[c]{@{}l@{}}- viewed as being outside of the ML workflow \\ \\ - solidified as axioms or hailed as requirements \\   or relegated as limitations\end{tabular} & \begin{tabular}[c]{@{}l@{}}- target of the assumption \\   often remains unstated\\ \\ - premise is seldom evaluated\end{tabular} \\ \cline{2-4}  
 & \begin{tabular}[c]{@{}l@{}}Relative\\ construction\end{tabular} & \begin{tabular}[c]{@{}l@{}} - defined in relation to data quality, model \\   specifications, or business objectives\\ \\ - rationalized in relation to ML workflow and \\   prevents deeper and more inclusive assessment\end{tabular} & \begin{tabular}[c]{@{}l@{}}- premise, target, and argument\\   are explicit and clear\\ \\ - identification and evaluation\\   of premises are easier\end{tabular} \\ \hline \hline
\multirow{2}{*}{\begin{tabular}[c]{@{}l@{}}Uncertainty about \\ what is being done \\ with assumptions\end{tabular}} & \begin{tabular}[c]{@{}l@{}}Integration with\\ existing workflow\end{tabular} & \begin{tabular}[c]{@{}l@{}}  \textbf{Reactive handling:}\\ - reactive and iterative approach to ML extends\\   to assumptions identification and handling\\ \\ - assumptions constructed relative to ML \\   workflows are often reactively handled\\ \\ \textbf{Unreflective quantification:}\\ - the incentive to quantify assumptions \\   evaluation obstructs reflective practice \\ \\ \textbf{Circle of ambiguity:}\\ - no mechanisms to capture evidence of \\   assumptions, creating more uncertainties\\ \\ - knowledge and communication gap in ML\\   between stakeholders creates circling ambiguity\end{tabular} & \begin{tabular}[c]{@{}l@{}}- formulation of argument's\\   target depends on how surprising \\ 
 assumption's  consequences are\\ \\ \\ - ambiguities in premises and \\   implied arguments are disguised\\ \\ \\ - target of one assumption forms\\   the premise of other targets\end{tabular} \\ \cline{2-4} 
 & \begin{tabular}[c]{@{}l@{}}Unstructured\\ documentation\end{tabular} & \begin{tabular}[c]{@{}l@{}}  \textbf{Informal and implicit recording:}\\ - distinction between formulation and installation\\   of assumptions is not clearly documented\\ \\ - site, content, and style of assumption recording\\   is strongly associated with role, resulting in conflict\\ \\ \textbf{Granularity of recording:}\\ - insistence to understand the rationale behind\\   assumptions lead to further assumptions based\\   on lived experience\\ \\ - no structured prompt to record the required level\\   of details in assumptions recording\end{tabular} & \begin{tabular}[c]{@{}l@{}}- lack of distinction between \\   formulation and installation \\   of premises in documentation\\   \\ \\ \\ \\ - premises are often recorded as \\   declarative statements with no\\   relation to target of the argument\end{tabular} \\ \hline
\end{tabular}
}
\caption{\rev{A summary of key themes, takeaways, and the premise-target theoretical lens we adopt to deconstruct and understand the confusing factors about assumptions in ML.}}
\label{tab:findings}
\end{table*}

\subsection {Ontological Differences}
\label{sec:ontology}

The conceptualization of an ``assumption'' varied greatly between participants. While some gravitated toward an understanding that coincided with their preexisting technical workflows, others perceived it as an isolated consideration, existing outside the normative bounds of development. These inconsistencies may be rooted in a systemic de-emphasis of abstract thinking, incentivizing participants to view socio-technical concepts as a static, external object rather than an embedded mentality \cite{selbst2019fairness,wang2022towards,malik2020hierarchy,fazelpour2020algorithmic}. When prompted explicitly to define ``assumptions'', many participants described it as preliminary, something to be ironed out, built upon, or invalidated. These initial assumptions also predicated on how future assumptions were handled. The elaboration of how this characterization plays out throughout the development process differed, with some participants conceiving of an assumption independent of internal components (section \ref{subsec:ind}) and others associating it directly or indirectly with other entities (section \ref{subsec:rel}). Both these constructions create uncertainties and confusion in their own ways, both in how they shape institutional handling of downstream tasks and potential harms. 
% Redundant:
% 
% Ultimately, it was the background and experience of the participant that dictated how they distinguished assumptions. 

\subsubsection{Independent Construction}
\label{subsec:ind}
\citet{delin1994assumption} describe how average discourse around an assumption implies that it is a sort of abstract entity existing in one's mind, and to interact with one is akin to finding, identifying, and examining a ``thing.'' This interpretation best characterizes what we label an \textit{independent} construction of an assumption. This type of assumption exists as an external \textit{other} to the primary subject---the ML workflow. The \textit{purity} of this workflow is a common theme among participants with a technical background. The idea of the model and the deference to ML work is a foundational mentality that conceptualizations of risk \cite{saxenaRethinkingRiskAlgorithmic2023,zanotti2024ai}, bias \cite{andrusWhatWeCan2021,kernAssumptionsBiasData}, and, in this context, assumptions must work around. 

This is best seen through how technologists frame assumptions as independent ideas that exist to serve the technology. For instance, for many participants, we introduced the concept of assumptions by inquiring the participant about \textit{givens}, or what information or knowledge the participant takes for granted. These givens are necessary \textit{premises} to begin a project as they unquestionably validate the primary heuristics of the project. In other words, the \textit{target} of independently constructed assumptions remains unstated or, in the best case, unscrutinized if it is explicitly mentioned. This is often because these assumptions lay the foundation of a project, and they are often immutable and the rest of the work must be accommodated. This immutability is justified as the organic nature of an AI model, with assumptions being reflective of its surrounding context rather than the technology itself. P1, an ML developer, explains the unspoken nature of these givens:

% Redundant:
% The emphasis of the technology as the central component of the product lifecycle further contributes to this mindset.
% P1, a technical product manager, elaborates on this:
% \begin{quote}
% \textit{"How I describe an assumption is something that’s taken for granted. Something where I don’t have to think about it. And it’s a fact or a basis for the work that I’m about to do. I don’t really need to question, hey, what’s happening? Is this what I need to do? It just happens, or it’s there, and there’s no question about the validity of it."} 
% \end{quote}
\begin{quote}
\textit{``I guess the most basic assumption is that the human behavior can be modeled with numbers...Because if you know these things are unquantifiable, then there’s no work...It's a set of axioms, I guess, from which I can draw conclusions. And if these axioms are violated, then, you know there’s no guarantee how the system will turn out.''} 
\end{quote}

% These givens are also provided through a proxy that is perceived trustworthy and thus rarely questioned. These may be institutionalized teams or roles that convey the givens directly or the background and experience of the technologist. 

Recent research suggests that the assumptions and choices of technical practitioners, in particular, are often found to be more subtle \cite{kery2019towards,kommiya2024towards,wang2019data}. These types of assumptions reflect a desired \textit{implicitness} to certain thinking that allows the technology to be developed in the first place. The assumption then possesses an \textit{interpretive flexibility} \cite{meyer2006three,star1989structure,leigh2010not}, being swept under the rug but still requiring further assumptions to validate it and allow for it to remain unspoken. This process may entail the transformation of an assumption into a \textit{requirement} or a \textit{limitation}. The former is managed through the validity of the assumption by an authority, which is either the assumer themselves and their expertise or a separate role or team that explicitly assigns it credence. They become \textit{informed} assumptions, legally justified and deliberated upon by personal decision-making. The latter may be designated as such through real-world constraints that prevent deeper internalization. Assumptions that are relegated to limitations may also be the byproduct of a ``perfection is the enemy of good'' culture \cite{sylvester2018applied,green2019good}. The assumptions that are not solidified as axioms, or hailed as requirements, or relegated as limitations, are then needed to be empirically validated and conceptually clarified in relation to the target towards which the assumptions are directed.
% in order to assess risk.

% Independent assumptions are treated as \textit{boundary objects} \cite{star1989structure,leigh2010not}, where the back-and-forth between ill-structured and well-structured forms are to be expected. 

% \begin{quote}
% \textit{I’m sure you know, the model would not get things correctly, like 100\% of the time. But you know, we don’t aim for perfection. We aim for some number like 99.9\% of the time. Maybe that’s good enough.} 
% \end{quote} (P2)
% \textit{…it feels like for a lot of product teams, they get hindered because of the years of risk. So they just really want to move forward so that they can actually deliver something.} 
% (P3)

\subsubsection{Relative Construction}
\label{subsec:rel}
Other participants had a more \textit{embedded} perspective on assumptions. While independent construction allowed the assumption to exist as its own object flowing through and being manipulated by the ML workflow, \textit{relative construction} implies that assumptions exist \textit{in relation} to existing phases of the development process. Returning to \citet{delin1994assumption}'s characterizations, this type of assumption involves examining the ``mental-event-or-state'' of an individual or institution instead of perceiving the assumption as an independent entity. In other words, relative assumptions exist as a byproduct of the practitioner's \textit{mentality} in a workflow rather than an external factor.

In particular, participants embodying this perspective defined assumptions through technical framing, citing decisions around data quality or model specifications. The assumptions must live within the confines of a technically-driven approach and be subject to dissection through that lens. This integration of assumptions into a technical dimension can help practitioners investigate deeper issues within their work, despite the purview being narrower. Since the target of the assumptions is often explicit in this case, identifying the argument is relatively easier than in independent construction. This has both upsides and downsides: it allows the practitioner to navigate the assumption through a familiar paradigm, but it also presents assumptions as an inevitability through the sheer breadth of available information. This could allow assumptions to go unchecked, but in a justified resignation, as demonstrated by P2, an ML scientist:

\begin{quote}
\textit{``...we kind of don't have the bandwidth to check each and everything. So that's maybe one assumption we make. We're also kind of assuming that...all of this leads to data correctness...but we end up making some assumptions like, for example, data is about accuracy. If data is empty, then you should just treat it as empty and not treat it, as you know, something meaningful...They are assumptions, and they will sort of, you know, go into the model and be baked into the process.''}
\end{quote}

% Relative assumptions, like independent assumptions, originate outside the realm of the ML workflow, but are \textit{actualized} through technical framing. 

Relative assumptions too possess an inherent implicitness that may unintentionally inform understanding of the model. But if an independent construction allows for an assumption to persist as its own object with the possibility of transforming into something beneficial to the technical process, then a relative construction attempts to integrate an assumption directly \textit{into} the process. The formulation of problems is an example, as they usually become intertwined with relative assumptions: P2 described how associations with specific demographics, for instance, may be less scrutinized due to the expectations of the technical team. In other words, if the data is labeled or categorized in a certain way that conforms to the lived experience of the practitioner, it may prevent deeper assessment. And because relative assumptions are tied directly to workflow components and workflow components are necessary, there is an incentive for the assumption to be rationalized.

% \textit{If you know, your assumptions start becoming givens, and you know ideally at the end of your project you should have only knowns and no assumptions, nothing left to assumption.} (P5)
% \\
% \textit{….because at the end of the day, bias in your model means bias in data and bias in data is just like reflecting what exists in the real world…So there those are certain assumptions in terms of like, 'oh, this bias already exists and you’re conforming to it in a certain sense…'} (P6)

Relative assumptions need not necessarily be attached to \textit{technical} processes only. A few practitioners in management roles described how assumptions can linger throughout business objectives and outcomes; what is deemed critical to operationalizing the company vision is often an assumption in and of itself. These assumptions are often second or third-order assumptions \cite{berman2001opening,korzybski1958science}, meaning that the assumer is multiple degrees removed from the original observation that incited the assumption\footnote{\rev{To understand the confusions around assumptions, for the sake of clarity, we almost exclusively treat an assumption independently of other assumptions. We leave explorations of higher order and connected assumptions to future work and briefly point to related works at the end of section \ref{disc:articulate}.}}. More so, the authority attached to the central teams that assert these claims makes it easier to internalize relative assumptions because the task of proving them right is often the primary function of the business.

% \begin{itemize}
%     \item 			i. Data collection quality - 30-35 age not checked - if model performs badly for this distribution, we won’t know
% \item			ii. Data correctness 
% 	\item		iii. Lack of data - loss to follow up before death - how to impute missing data
% 	\item		iv. Data quality is often equated to data accuracy
% \item First data quality is investigated because that is often taken for granted
% \end{itemize}
% -  Investigate assumptions if personally relevant - else take altruistic stance

% <1 para>
% - Another conception of assumptions has an implicit relation with limitations
% - though when practitioners talk about limitations, assumptions are inherently present, they are unable to describe it explicitly.
% - Unable to describe assumptions as limitations - the latter is always in terms of outcomes - except a few - Assumptions are ignorant decision or requirements - can be seen as limitations
% - Though assumptions inform various actions/decisions, focus is on latter and not on former

% <unused for now>
% - Assumptions in operationalizing business vision - economic impact vs revenue example - 40\% assumptions comes from objective or mission statements - intentionally used - anticipate

\subsection {Procedural Uncertainties}
\label{sec:procedure}

Other than ontological differences, practitioners also face several challenges in determining what to do with the assumptions. In this section, we lay out the uncertainties that accompany various \textit{procedures} practitioners employ to identify and handle assumptions. We find that these uncertainties either correspond to integration methods with existing workflows (section \ref{subsec:integrate}) or the documentation practices for collaboration and accountability (section \ref{subsec:doc}). 

% Finally, we discuss the strategies practitioners employ, and the challenges they face to bring some structure to the articulation of assumptions for alleviating some of the surrounding confusion (section \ref{sec:articulate}.) 

\subsubsection{Integration with existing workflow}
\label{subsec:integrate}
\citet{brookfield1992uncovering} argues that the investigation of assumptions must be a deliberate and reflective process to assess and validate various decisions that go into a process. However, most practitioners we spoke to shared that their typical ML lifecycles in practice seldom offer them avenues to reflectively identify and handle assumptions.

\smallskip
\noindent \textbf{Reactive Handling.} Machine learning in practice is usually an iterative and outcome-oriented pursuit \cite{ashmore2021assuring,malik2020hierarchy}: almost all our participants start their ML lifecycles with lesser information than they ideally need, but then iterate and learn about assumptions as they go. Because there is no appropriate structure for them to inquire about assumptions, this iterative process often puts practitioners in ambiguous situations where there are no clear and accepted procedures to follow when one is found. This, in turn, demands practitioners to adopt a \textit{reactive} approach to identify and handle assumptions. This is most prominent in relative assumptions, in which the assumption is directly embedded within the ML workflow. For instance, some practitioners, such as P4, raise a red flag and look for assumptions in data when presented with a ``neat classification problem'' in which all data is categorized too perfectly. For a few others, assumptions are investigated only during exploratory data analysis (EDA for short) when extreme and skewed patterns about data are observed. \rev{P3, a technical lead, provides an example of this mentality by demonstrating how their team took certain factors for granted until a ``surprising'' result was found:}
\begin{quote}
    \textit{\rev{``So in any modeling process, we do some basic EDA to understand the quality of the data. But some things we take for granted. We then build a model. And sometimes, surprisingly, results will be like, very good than what you anticipate. Then we start looking into the top features. And then we think about, okay, is there any data leakage? This is very hard to understand when you're doing initial EDAs, right? But once we build the model, once you're seeing surprising results, then we go back to business. Then we discuss its implications...is there anything wrong with that feature?''}}
\end{quote}

Because relative assumptions can be associated with both technical and business processes, what is deemed surprising by the latter may inform the reaction of the former. This is demonstrated by P5, a lead data scientist, who articulated how the inquiry of assumptions often takes place only when the outcomes have unanticipated business implications in the typical ML workflow.
While the above-discussed ``innocent until proven guilty mindset'' (P6) is prevalent among ML teams, some practitioners in management-related roles highlighted that product teams generally use existing collaboration infrastructure, such as GitLab\footnote{https://about.gitlab.com/}, Asana\footnote{https://asana.com/}, or other business management software to inadvertently integrate assumptions discussions into the typical workflow. Though not a typical practice of ML teams, a couple of practitioners also shared that dedicated ``assumptions trackers'' are sometimes used to encourage documentation and discussion of assumptions at work. 

However, P6 shared that the use of these tools is largely descriptive and only results in noting down the simplistic description of assumptions, subjective association with the argument's target, and mapped ownership of assumptions redressal. 
Overall, assumptions are communicated through these tools, but it is unclear if they are reasoned and rationalized: questions such as why the assumptions are made, why they are necessary, what kind of assumptions they are, what their implications for ML lifecycle are, etc. are sparsely addressed by these tools. We expand upon this tendency to examine assumptions only at a surface-level in the next section.
% quote : i have a decade long experience in client facing roles - with great confidence i can say that this is not the case

% - though assumptions tracker etc. could be helpful, it often requires a historian type method - As historian, methodology revolves around recognition of “contingency” - reflective endorsement
% -- case study analysis

% Finally, integration into existing workflow also is contingent on the cost of 
% <1 para> - hold this for now
% With recent increase in RML attention, there has been some efforts to invoke assumptions - however
% 1. RML docs are not universal - best used for system documentations
% 3. RML toolkits are not used in many organizations 
%     a. Assumptions work comes at a cost - RML doesn’t help - Mediterranean addition example
% 4. RML just borrows inputs from various laws - we need dedicated legal frameworks.

\smallskip
\noindent \textbf{Unreflective Practice through Quantification.} A crucial constituent of assumptions identification and assessment is the \textit{reflectiveness}, if not \textit{critical reflectiveness}, of the process \cite{paul1993critical,paul1993workshop}. 
However, \textit{the seductions of quantification} \cite{merry2009seductions} often derail practitioners from performing reflective exercises on assumptions and instead reroute them to aspire for a fictitious objective and unequivocal state. \rev{P10, an engineering consultant, elaborated on her perceived futility in quantifying assumptions:}

\begin{quote}
   \rev{ \textit{``...we do have some in-house metrics…we also really want to know which task is being performed the best without any sort of assumptions or biases so we have inbuilt like a comparator for ourselves, based on particular metrics… I believe these metrics can be very much subjective, based on the use case a particular company or a particular product has...there is no way we can quantify what assumptions a particular model is taking…''}
}
\end{quote}

\textit{Independent} assumptions are most prone to this instinct---they must adhere to the workflow. Practitioners noted that the organizational constraints and standard ML workflow practices do not incentivize many technical counterparts to perform reflective exercises but only make them care about data coverage and infrastructure-related concerns. But this interaction between assumption and workflow is bi-directional. Both in their own practice and during our case analyses, practitioners suggested ways of using various computational methods \cite{yanga2024exploratory,zhang2018empirical} to debug and act on their assumptions. However, as an extreme example, one practitioner supported the use of software plugin-type tools to automatically check all assumptions and provide a score to quantify assumptions evaluation. 

In our case analyses with practitioners, we find that many of them spend more time understanding and justifying various computational steps in LLM reports but focus less on reflecting on the premises and conclusions that inform the various steps they are analyzing. We observe that practitioners with some exposure to responsible ML principles often recognize that over-reliance on computational methods just disguises the underlying ambiguities, not removes them \cite{green2018myth,fazelpour2020algorithmic}. This means that the organizational instinct and incentive to adhere to the ML workflow implies a bias to a relative construction of assumptions. They also shared that such disguising has the danger of pushing practitioners to ignore the assumptions and the implied argument, which often peek through at later phases of ML lifecycles in various forms. \rev{P7, a designer working with state organizations, shared how a risk assessment tool unnecessarily quantified risk evaluation:}

\begin{quote}
    \textit{\rev{``so these risk assessment questionnaires are designed such that you're kind of casting a wide net like, yeah, you're probably at risk of something. After they do the risk assessment, the general next step is, your team has to come up with a control plan in detail explaining how you're going to mitigate those risks. So you know, obviously, some of the feedback that we got was like, okay, so we could come up with the most horrific AI system that has all these risks to end users. And it's all okay, because we can just write a control plan, saying that we'll mitigate right? So there was a lot of like contention with that approach.''}}
\end{quote}

While the quantification of assumptions is intended to reduce uncertainty around decision-making, further assumptions emerge when model performance is negotiated with other sociotechnical concerns, such as safety and fairness. For many practitioners, this leads to a nuanced tension they face when, for instance, deciding on a lower threshold for the maximum number of security violations. For example, while a 90\% violation is clearly red-flagged, negotiation between 0.5\% and 1.0\% creates debates and confusion. We expand on the cyclical nature of assumptions in the next section.

% benchmarks curb reflectivity?
% <-- finally benchamarks are blindly used (cite gorilla etc.) - see quote about assumptions>

% - it was not a one side affair - some practiotners also sggested ways of reflectively using compuational methods - some also did it in our case studies
% -- while some computational techniques aid assumptiosn inquirt reflective - CF and knowledge base
% - however, the infrastructure (connect to worlflow + documnet next)

% -- embarassement score    

% When participants were asked to identify assumptions more indirectly through analyzing model documentations at face-value, some associated anything distinct from technical terminology as an assumption.

\smallskip
\noindent \textbf{The Circle of Ambiguity.}
While assumptions are often invoked to mitigate the ambiguity around unresolved questions, ironically, due to their very implicit nature, they instead raise several questions and concerns they are trying to ameliorate in the first place. Consider the use of automated risk-assessment tools used in many organizations in recent years: practitioners find that many responses to assessment software require extrapolation and reflection. However, the checklist-type design and quantification of risk evaluation in these tools obscure the \textit{consequences} of underlying assumptions. A few participants shared that several sub-modules of these tools in their organization require practitioners to invoke assumptions in their own responses, but there are typically no mechanisms to capture the evidence of these assumptions, creating more uncertainties. Therefore, the trajectory of an assumption in an ML workflow is one that may beget more assumptions. This ambiguity is discussed in \textit{Critical Thinking} as complex or \textit{chain of reasoning} arguments, where the assumption that supports a target may be the target of one or more assumptions, and so on indefinitely \cite{hitchcock2021concept}. 

% Looking through the analytical lens our practitioners took in the case studies, our findings highlight nuanced uncertainties that motivate technical practitioners to make subtle assumptions.  

Consider P4, an ML scientist, who expresses how the confusion around calibration scores \cite{pleiss2017fairness} among different non-technical stakeholders motivated their subjective decision on how to present the model outcomes:

\begin{quote}
    \textit{``...for multiple projects that I was on...I saw that the easiest thing to do was to provide high, medium, low kind of bucketing of confidences rather than to provide a confidence score. The assumption that kind of affords a user is that both significant digits are significant, that there is a difference between, you know, 0.95 and 0.97 or something like that. And that really wasn't the case. As a practitioner, I would see that and just be like, okay, that's not that significant. But how does especially someone who is not used to reasoning about the processes that produce these numbers know that?''
}
\end{quote}

The ordinal categories expected to resolve misinterpretation of double-precision metrics yield further confusion for both technical and non-technical stakeholders depending on how risk levels are construed. A possible explanation for this circling ambiguity, as indicated by some of our participants, could be attributed to the knowledge and communication gaps between people with and without ML backgrounds \cite{chen2021beyond,kommiya2024towards,varanasi2023currently}. P7 shared that their ML scientists \textit{``did not feel like there was a lot of risk involved''} when the models they work with are more interpretable to them, in contrast to what risk-averse employees felt about the model and its policy implications. This logic informs another assumption: to fix biases, what is needed is \textit{more} knowledge and training. However, most management-related practitioners in our sample concurred that technical knowledge in their organization is generally very isolated and emphasized the need to document differences in interpretations and motivating assumptions.

% Because products are made in the perspective of the organization, they can be easily dismissed to fall under the responsibility of a different position.

% \begin{quote}
%     \textit{At least in my experience, providing confidence scores, was not always a benefit, because the ways in which users interpreted those confidence scores varied a lot right. Some people would really set a lot of store by the difference between 0.90 and 0.92, and not care about the difference between 0.9 and 0.7. And so for multiple projects that I was on you know, I saw that the the easiest thing to do was to provide high, medium, low kind of like bucketing of confidences rather than to provide a confidence score. The assumption that kind of affords a user is that both significant digits are significant, that there is a difference between, you know, 0.95 and 0.97 or something like that. And that really like that wasn't the case. As a practitioner, I would see that and just be like, Okay, that's not that significant. But how do especially someone who is not used to reasoning about the processes that produce these numbers know that?
% }
% \end{quote}
% - exclusion complex
% - Some organizations have top-down assumptions handling 
% - 

% - we found similar observations - Mihir doc - quote - assumptins  techincal - infor  subjective deicon - rationale lost

\subsubsection {Unstructured Documentation}
\label{subsec:doc}

Though the successful adoption of responsible ML principles in organizational settings is open to debate \cite{deshpande2022responsible,rakova2021responsible,raiAdoption}, practitioners are undoubtedly getting increased exposure to different toolkits and frameworks for responsible ML \cite{liang2024systematic,yang2024navigating}. While most of these support systems have some reference for practitioners to elicit and discuss assumptions behind various decisions (section \ref{rel:periphery}), prior works are unclear on how and to what extent assumptions are documented through these toolkits from the perspective of a practitioner. Below, we discuss two characteristics of documentation practices that contribute to the confusion around assumptions.
% For instance, a recent study found that the use of framework such as modelcards have rapidly risen in the last few years.

\smallskip
\noindent \textbf{Informal and Implicit Recording.} \citet{delin1994assumption}, in their discussion about assumptions, make a distinction between formulating and installing an assumption: while \textit{formulating} is about the expression of intent to install an assumption, \textit{installation} corresponds to aligning the execution to the intention within given constraints. In one of the excerpts in our case study (appendix \ref{appendix} \cite[p.~64]{anil2023palm}), the authors use ``marked references'' or annotated characterizations of identity\footnote{Some participants speculated the assumptions behind the usage of this term, but for this discussion, this can be perceived as typical annotations in ML data pipelines.} to assess the representational bias of their language model. Though our participants did not use the terms referenced in the \textit{Informal Logic} literature exactly, they did observe that if the authors had intended to operationalize representativeness with these annotations, then they must have correctly \textit{installed} the assumptions but did not explicitly document the \textit{formulation}. While the authors of the case study vaguely note down this assumption in the limitations section of the report, many of our participants were dissatisfied and shared that assumptions that are not articulated at the time of their identification get forgotten over time. 

Because the recording of assumptions is often informal and unstructured, system documentation and other related trackers are rarely modified after an assumption is acted upon. While some practitioners stated that their organizations require them to record assumptions in separate sections in accordance with a design or product requirement, others just recorded them offhand in change logs. Both approaches offered no structured prompt or instructions about how to aptly record them. As such, many participants hinted that documentation is not often given serious thought for internal projects. Below, P9, another technical lead, admits how these details are either not documented explicitly or get lost in an attempt to simplify the language.

\begin{quote}
\textit{``...when I run into an assumption like that, I try to distill it sometimes through multiple rewrites into a simple, concise, clear declaratory statement which can be connected to others...I would forget what I had in mind when I was writing those things down...'' 
}
\end{quote} 

The site, content, and style of assumptions documentation also have a strong relation with the primary role and responsibility of the practitioner, leading to conflicting situations when one party (say an ML scientist) has to consume and work with assumptions recorded by another party (say a product manager). For instance, writing and analyzing compliance-related reports are often perceived as the responsibility of safety or legal teams. Their articulation of assumptions might differ from that of developers and data scientists, who usually write in terms of inputs and outputs. If these technical practitioners are tasked to write about assumptions, they are less likely to appreciate the implications of their assumptions and choices in their reports. As the data scientist P5 mentioned, when they use LLMs, they go directly to model docs, datasets, and benchmarks, and treat the system cards and other reports accompanying these LLMs (with safety and RAI analysis) as just \textit{``terms and conditions.''} This fundamental asymmetry presents a fragmentation of thinking and recording that has consequences for how assumptions are handled. 

% 		ii. RML just borrows inputs from various laws - we need dedicated legal frameworks.
% Some implicit decisions require technical effort - need better documentations

% In this section, we will use an example from our case study where the LLM authors mention that they \textit{``excluded data from certain sites known to contain a high volume of personal information about private individuals''} (cite).

\smallskip 
\noindent \textbf{Granularity of recording.} We leverage the \textit{analytical} perspective our practitioners took in the case studies in assessing how deliberate they were in recording assumptions. As \citet{brookfield1992uncovering} argues, this particular type of perspective is well-positioned to step out of a familiar interpretive frame of reference and look at assumptions through an unfamiliar lens. We observe that the most common practice the participants followed is simply listing down the identified assumptions in some style and form, though at varied levels of formality, such as they would in a business requirement document or vocally at a team meeting. However, practitioners note that this method generally relies on their own unjustified claims. In other words, there was no distinction between the premises, target, and the argument that the assumption-target complex makes.

While listing down assumptions found in the case study, most of the practitioners sought further clarification on \textit{why} particular assumptions were made. This phase is when the interpretive lens of a practitioner begins to crack, and their primary disciplinary training starts to dominate; the choice of assumption to expand on and what needs to be explained starts to depend again on their organizational role or lived experience. For instance, when technical practitioners observe relative qualifiers such as ``higher'' or ``lower'', their scrutiny often stops at the sight of a quantified relation, such as ``40\% higher.'' \rev{Participants shared that many technical practitioners} end up not reflecting on how benchmarks are only the \textit{indications} and not the equivalence of the capability they are measuring. \rev{In the words of P11, an AI ethicist:}

\begin{quote}
    \textit{\rev{``The fact that an LLM could perform well on an LSAT benchmark means not necessarily that it's capable of, you know, of solving legal problems. But it could be quite capable of delivering to you the outputs that mimic responses to those questions and that's valuable. Where we can go wrong is to infer that, you know, because the model can pass a test that must mean that its feature representations allow it to understand the material of the test. This is a very different question and that I feel like needs holistic evaluations. You know, once they're developed, there's a huge evaluation gap here.''}}
\end{quote}

% \rev{Below, P4 explains how the granularity of assumptions inquiry ends when benchmarks are conflated with the capability they are intended to measure:}

% \begin{quote}
%     \textit{\rev{``if your model can solve this entailment data set, the idea is that it can do textual entailment more broadly. But there are several kinds of subpopulations that you need to reason about before you can actually take that to the task that you actually care about. It's reasonable with the intentional definition of entailment that is presented to say, like, oh, my model can understand that 3 tenths and 30\% of the same thing. That is like a linguistic equivalence. But models don't do that right. So if you are building an entailment based application that relies on that kind of numerical reasoning, your application will break. That's not an assumption that you can actually...do textual entailment as a capability rather than textual entailment on MultiNLI as a data set.''}}
% \end{quote}

How a clarification is or ought to be justified is a critical step in assumption recording, the complexities of which are expanded upon in the following section. P8, an AI governance architect, makes a distinction between \textit{visibility} and \textit{explainability} in recording assumptions to avoid confusion. While the former could be one or more declarative statements that add a marginal level of detail, the latter is about reasoning in every step with logical validity. For instance, in many of the LLM reports, our participants observed that data quality is justified by arguing that the followed data processing steps led to ``good'' model outcome in terms of some metrics. However, the above aligns more with the visibility-level of clarification. An explainability-level of granularity would involve expanding how quality is operationalized (for example, showing no outliers, following a representative distribution, etc.) and then logically explaining how the data can be assessed on both these parameters and the model outcome.

Finally, the maximum level of granularity in which an assumption can be recorded is ambiguous and context-dependent. While some practitioners seek ``expert'' intervention, which entails transferring the authority to a team lead or an executive, most practitioners instead advocate for a diverse collaborative effort to avoid biases. In section 5.1, we discuss a framework inspired by argument structuring in Informal Logic to articulate assumptions that facilitate such discussions.

% To do:
% 1. change quotes repetitive participants
% 2. include citations
% 3. check tense consistency

% We observe that practitioners implicitly or explicitly follow three distinct levels of granularity in recording assumptions. 

%% file: 5.discussion.tex
\section{Discussion and Future Work}
\label{sec:discussion}

We highlight throughout our findings that since arguments need not be explicit statements expressed in documentations, there is an inherent subjectivity in interpreting and analyzing assumptions in arguments, causing confusion when practitioners work with different stakeholders. Hence, adequately recording the assumptions and clearly situating them in arguments is essential to work through subjective interpretations. To motivate this line of work, in section \ref{disc:articulate}, by connecting the premise-target lens with our empirical findings, we derive a framework to \textit{articulate} assumptions for analysis. Then in section \ref{disc:creative}, we discuss how critical assessment of assumptions is also a lateral and creative activity \cite{fisher1997critical,delin1994assumption}. More broadly, our empirical findings and theoretical grounding lay out the foundation to strengthen the infrastructure for assumptions identification and handling in machine learning.

% As \citet{oulasvirta2016hci} argues, without strong conceptual understanding of how a problem is situated, 
% - cehck future roadmap last section:
%     - without conceptual rigor construcute solutions cannot happen - link both types - out wrosk contributes
% - we offer one example fo construcutvt snapshot - inferred form our findings - a frameworrk to articulate assumptiosn -t the first step - 
%     - see below point
% - this can also be assessed as improving other characteristc such as efficacy which we did not do with our findings and current RQ 
% - but we don't completely agree with this due to boudary objects which we discuss in 5.2, and discuss certain fuure directions inspired from our findings, grounding in theory

% -- significance: important to practtioers
% - what can they now acheive nefore reading the paper
%     - significance for them of the problem at hand
% - table - importance of the i m- provement for stakeholders
%     - capacity

\subsection{Explicating the Articulation of Assumptions}
\label{disc:articulate}

We discuss in section \ref{subsec:doc} that the documentation practice of assumptions is usually implicit and unstructured: it can be in the form of simply listing them in varied styles, inline recording their presence in change logs, or including/excluding justifications of claims. So when a practitioner attempts to analyze either their or another person's assumptions at a later point in time, they could have several unresolved questions and doubts, as evidenced throughout our findings. Further, how a practitioner conceptualizes an assumption, either as an independent axiom or in relation to existing phases of the development process (section \ref{sec:ontology}), influences their course of action in the ML workflow.
Therefore, explicating the articulation of an assumption is an essential ingredient to clearing out many surrounding confusions.

It should also be noted that explicating an assumption, in essence, implies clarifying or detailing the broader argument in which that assumption is embedded. In other words, in articulating an assumption (and thereby an argument), a clear distinction must be made between the premise (i.e., the assumption) and the target the premises are supporting to make an argument. Prior works argue that presenting an argument in this premise-target frame is one of the most effective strategies to communicate the proposal the argument is trying to make, where ``effective'' signifies the increase in likelihood of scrutiny from the audience of the argument \cite{innocenti2021constructing,kauffeld1998presumptions,jacobs2000rhetoric}.

Now, our findings describe that some practitioners attempt to ``justify'' their use of assumptions in their documentations (section \ref{subsec:doc}).
However, this reference to justification creates different interpretations of assumptions where the distinction is not always clear.
For instance, does justification imply the act of using assumptions for achieving something or refer to the conclusion the assumption is trying to support?
In philosophical terms, the illocutionary acts of justification could vary depending on the context and create more uncertainties\footnote{We discuss in section \ref{rel:core} that a premise is composed of the proposition and the illocutionary act (or the accompanying intentional state) but do not engage with this decomposition in our findings. We leave further exploration to future works.}.
In contrast, in articulating assumptions as premises to a target, the \textit{object} of scrutiny becomes much clearer. So, this premise-target complex becomes the \textit{core} of our articulation framework (Table \ref{tab:explicate}).

Our findings also discuss that practitioners can formulate different assumptions for the same target, depending on their conceptualization of assumptions, organizational constraints, or disciplinary backgrounds.
To accommodate such differences within the premise-target frame, we add a meta-layer to the explication---which we call the \textit{differentiator}---that distinguishes three broad categories in which the premises can be related to the target. 
We adapt the assumptional categories discussed by \citet{brookfield1995getting} to our findings and propose three types of differentiators that cluster different assumptions.
% as a function of state and action. 
First, assumptions can be about the structural \textit{understanding} of different elements of the target, such as a sociotechnical idea that underpins the target.
In the second category, the assumptions are about the \textit{appropriateness of the action(s)} performed to achieve the target. 
The final category is about how the performed action will \textit{address} the target.  
The framework also requires mentioning if the target and assumption are implicit. 

To illustrate, let's analyze the below excerpt from the technical report of a large language model, PaLM 2 \cite[p.~87]{anil2023palm}. 
\begin{quote}
    \textit{Are there certain annotator perspectives or subgroups whose participation was prioritized? If so, how were these perspectives sought out? \textbf{No.}}
\end{quote}
The object of scrutiny here is the response provided by the report's authors for a CrowdWorksheets question on how annotators were chosen \cite{diaz2022crowdworksheets}. 
We chose this example as many practitioners could (and did in our case studies) straightforwardly recognize the existence of assumptions within it.
\rev{Such responses also reflect how practitioners pay little attention when filling out details on biases or limitations in popularly used toolkits such as model cards \cite{liang2024systematic}.}

Table \ref{tab:explicate} presents our analysis.
Now, depending on the differentiator lens a practitioner takes, different assumptions can be articulated for the same identified target.
The first assumption from the left is about the assumer's understanding of the concepts within the target: the inclusivity of the annotator group.
The second category of articulation is about why the alternative course of action, such as leveling up or down the representation of a group \cite{mittelstadt2023unfairness,corbett-daviesMeasureMismeasureFairness2018}, is inferior to taking no action to achieve the target.
The last type of assumption discusses how when no group is explicitly prioritized, the hiring of annotators stays inclusive.
The assumptions in different categories can also relate to one another. 
For instance, in this case, the first assumption on understanding inclusion could inform the third assumption on how performing no action will address the target objective.
Finally, while we have discussed only a single assumption per category, in other situations, multiple assumptions may also be grouped just under one category.

\rev{In general, while commonly used toolkits simply prompt practitioners to list or explain why something is an assumption (section \ref{rel:periphery}), our framework presents a method to specifically break down the articulation of an assumption by considering the factors rooted in confusion we discussed in our findings. It is important to note that the tabular format we have used here is to illustrate the building blocks of \textit{one} assumption. However, in practice, we may want to articulate more complex and dynamic relations between assumptions in general, such as how the understanding of one assumption changes with another or how addressal of some assumptions updates others. Several works in Informal Logic have discussed the benefits of software programs and visual markers to interactively map and understand such changes in premises and targets \cite{daviesusing,harrell2008no,martin2009computer}. We encourage the HCI and responsible ML community to build on these works to logically understand the flow of assumptions.} 

% Though there is no tool such as an ``assumptions sheet,'' 

% -- used assumption is always about absene than about rpesne - discussion
% -- Another example is indigenous - explicating articulation discussion

\subsection{Assumptions Inquiry as a Lateral Thinking Process}
\label{disc:creative}

\begin{table*}[t]
\centering
\caption{\textbf{Articulation of \rev{an Assumption} in an Argument.} We offer a framework that breaks down the articulation of assumption through three differentiators: the \textit{understanding of the target}, the \textit{appropriateness of performed action}, and the \textit{addressal of the target}. These dimensions convey the primary components of a premise-target argument and how an assumption can be embedded within it.}
\resizebox{0.9\textwidth}{!}{%
\begin{tabular}{|llll|}
                          \hline
                          \multicolumn{4}{|c|}{}\\
                          & \multicolumn{1}{r}{\textbf{Target:}}                                                                                                                    & To perform inclusive hiring of annotators                                                                                                                       &                                                                                                         \\
                                                                             & \multicolumn{1}{r}{\textbf{Is Target Implicit?:}}                                                                                                       & No                                                                                                                                                              &                                                                                                         \\
 \multicolumn{4}{|c|}{}\\ \hline
                                                                             \multicolumn{4}{|c|}{}\\
\multicolumn{1}{|c}{\textbf{Differentiator:}}                                 & \multicolumn{1}{c}{\textit{Understanding of the target}}                                                                                                    & \multicolumn{1}{c}{\textit{Appropriateness of performed action}}                                                                                                & \multicolumn{1}{c|}{\textit{Addressal of the target}}                                                        \\
 \multicolumn{4}{|c|}{}\\ \hline
& & & \\
\multicolumn{1}{|c}{\textbf{Assumption:}}                                     & \begin{tabular}[c]{@{}l@{}}The default state of the world \\ has equal representation. \\ So there is no need to \\ prioritize any group.\end{tabular} & \begin{tabular}[c]{@{}l@{}}Of several actions that can be performed, \\ such as leveling up/down representation \\ of a group, doing nothing is the most \\ appropriate action.\end{tabular} & \begin{tabular}[c|]{@{}l@{}}When no group is explicitly\\ prioritized, hiring is inclusive. \end{tabular} \\
 \multicolumn{4}{|c|}{}\\ \hline
\multicolumn{4}{|c|}{}\\
\textbf{\begin{tabular}[c]{@{}c@{}}Is Assumption \\ Implicit?:\end{tabular}} & Yes                                                                                                                                                     & Yes                                                                                                                                                             & Yes                                                                                         \\ 
\multicolumn{4}{|c|}{}\\
\hline           
\end{tabular}
}
\label{tab:explicate}
\end{table*}

Section \ref{sec:procedure} discusses how, despite the rise in responsible AI toolkits, articulation of assumptions in practice is often informal and unreflective. 
However, the exercise in Table \ref{tab:explicate} suggests that the articulation of assumptions can involve deliberate effort and reflective thinking on how premises are connected to the target, and decides future course of action.
For instance, for the above example, a computationally-oriented practitioner might think about assumptions in terms of algorithmic fairness and analyze how performing no action provides more ``fair'' hiring outcomes than leveling down or up the fairness scores of particular sub-groups \cite{mittelstadt2023unfairness,corbett-daviesMeasureMismeasureFairness2018,cooper2020normative}.
Others might focus on the assumptions in the understanding of sociotechnical factors, such as inclusion or fairness, to analyze how it affects hiring and various decisions in the workflow \cite{wangFactorsInfluencingPerceived2020,pfeiffer2023algorithmic,dolata2022sociotechnical}.
Our framework simplifies this articulation process for practitioners in communicating their argument and supports them in deciding future courses of action while also revealing how articulation in an assumption inquiry requires practitioners to think more proactively about how and why an assumption is made.

Assumptions inquiry is \rev{broader and more} complicated than \textit{only} articulating. Our articulation framework in Table \ref{tab:explicate} presupposes that assumptions are already identified. While the premise-target-differentiator complex can help surface assumptions, for many cases, the assumptions identification process itself requires employing systematic exercise and techniques \cite{walton2012argument,palau2009argumentation}, as illustrated in our example. We discuss in section \ref{subsec:doc} that even for practitioners employing responsible AI toolkits to track down what they think are assumptions, there is little support provided to carefully reflect on assumptions; for instance, when they may inquire why they think something is an assumption or analyze what an assumption entails. Though our articulation framework sets up the stage for that reflection and subsequent course of action, practitioners may still face several questions on how to evaluate or respond to assumptions and arguments. We call the attention of HCI and responsible AI researchers to the analytical methods offered by related works in Informal Logic to systematically perform these functions \cite{walton2008argumentation,govier2018problems,govier2010practical,blair2019judging}. These works suggest that thinking about assumptions involving several non-sequential overlapping processes---such as identification, articulation, evaluating, and responding---demands lateral thinking and needs to be separated from the activities of a typical workflow. In other words, the \textit{thinking} process around assumptions must continue in parallel to the ML workflow.
 
Overall, our findings highlight an inadequacy in existing infrastructures of ML ecosystems in thinking about assumptions and discussing the contributing factors of the confusion resulting from this inadequacy. 
However, instead of focusing on the development of more toolkits exclusively for assumptions, we recommend practitioners to be more deliberate in their assumptions inquiry using the toolkits at their disposal. Prior works have argued against the techno-solutionist adoption of these toolkits where the mere \textit{following} of specific processes is often assumed to solve ethical or fairness issues \cite{wong2023seeing,balayn2023fairness,dengExploringHowMachine2022,lee2021landscape}. But this practice continues to exist and is reflected in how the technical reports and documentation of large language models are currently drafted, such as in the responses to model cards \cite{mitchellModelCardsModel2019} or datasheets \cite{gebruDatasheetsDatasets2021} where there is less clarity on the reflectivity that goes into this process \cite{wong2023seeing,madaioCoDesigningChecklistsUnderstand2020,crisan2022interactive,bhat2023aspirations}. The example we discussed in the previous section depicts this superficiality. Further, the assessment of these toolkits too, in practice, predominantly focuses  on the \textit{completion} of different sections of these toolkits \cite{liang2024systematic,yang2024navigating}.

This trend is particularly challenging for assumptions inquiry because, by definition, the steps that go into assumptions identification, articulation, and evaluation are not something that can be merely followed through instructions but rather need to have deeper engagement.
Similarly, the assessment process is further complicated because the analyst is now interpreting someone else's assumptions inquiry, which cannot be just evaluated based on rubrics or completeness \cite{govier2010practical,blair2019judging}. 
Having said that, we do not imply that toolkits should not be developed to bring structure to the assumptions inquiry process; in fact, \rev{as a first step,} we offered a framework to articulate assumptions in the previous section that \rev{future works can integrate} within existing or new toolkits. 
\rev{Prior works in Informal Logic appreciate the complexity of assessing arguments and their premises by considering several factors such as soundness, acceptability, relevance, sufficiency, and fallacies \cite{blair2019judging,govier2018problems}. Evaluating our articulation framework along these and other relevant dimensions to ML is an interesting area to explore in the future.}

\rev{Finally, it is important to note that we laid out our articulation framework from the perspective of an assumption \textit{analyst} who may use this framework in (in)formal discussions, business management platforms, quality assessment phases, or throughout the ML lifecycle. However, to guide an \textit{assumer} who wants to demonstrate accountability in their decisions, further research is required to assess how effectively constituents of an assumption can be presented, for instance, by integrating it with particular modules of existing toolkits and frameworks, or by consolidating all assumption break-downs in a dedicated place in the form of an ``assumptions sheet.''}

\section{Conclusion}
\rev{In this work, we explored an overlooked phenomenon---confusions around assumptions in machine learning---and investigated the factors contributing to this confusion.}
\rev{We ground our empirical contributions in the argumentation theory of Informal Logic to explain the nagging confusions surrounding assumptions in ML. In particular, by viewing assumptions through the premise-target lens of an argument, our findings throw light on why and how assumptions are made in ML and support practitioners in understanding and resolving confusions around assumptions.}
While we deconstruct these confusions that are peripherally discussed in prior works, more uncertainties still remain on the adoption of assumptions inquiry in practice. We invite other researchers to build off of our work and reflect critically on their \textit{own} assumptions in how they imagine and craft new ethical or fairness toolkits---not only as a function of data or model requirements or in the form of checklists and questionnaires, but through the logic and thinking of practitioners in pragmatic contexts.

%% file: 6.appendix.tex
% \newpage

\appendix

\section{Appendix}

\subsection{Case Study Samples}
\label{appendix}
The following are paraphrased samples of text extracted from various LLM documentations. They were employed during the second phase of our remote interviews with participants, in which they vocally reflected on each statement and considered possible assumptions.

\smallskip
\noindent \textbf{PaLM 2 \cite[p.~64]{anil2023palm}.} \textit{We see major differences in observed representations of people within pre-training data across all dimensions we consider. For sexuality, we see that while marked references are only found in 3\% of documents, most references are related to ``gay'' (53\%) and ``homosexuality'' (22\%)}.

\smallskip
\noindent \textbf{BLOOM \cite[p.~41]{le2023bloom}.} \textit{The library relies on minimal pairs to compare a stereotyped statement and a non-stereotyped statement (e.g. ``Women can’t drive.'' is a gender stereotype while ``Men can’t drive'' is not). The two statements differ only by the social category targeted by the stereotype and that social category is present in the stereotyped statement and not in the non-stereotyped statement. The evaluation aims at assessing systematic preference of models for stereotyped statements.}

\smallskip
\noindent \textbf{LLAMA 2 \cite[p.~20]{touvron2023llama}.} \textit{We followed Meta’s standard privacy and legal review processes for each dataset used in training...We excluded data from certain sites known to contain a high volume of personal information about private individuals...Sharing our models broadly will reduce the need for others to train similar models. No additional filtering was conducted on the datasets, to allow the LLM to be more widely usable across tasks (e.g., it can be better used for hate speech classification), while avoiding the potential for the accidental demographic erasure sometimes caused by over-scrubbing.}

\smallskip
\noindent \textbf{PaLM 2 \cite[p.~20]{anil2023palm}.} \textit{We additionally analyze potential toxic language harms across languages, datasets, and prompts referencing identity terms...Similarly, when disaggregating by identity term we find biases in how potential toxic language harm vary across language. For instance, queries referencing the ``Black'' and ``White'' identity group lead to higher toxicity rates in English, German and Portuguese compared to other languages, and queries referencing ``Judaism'' and ``Islam'' produce toxic responses more often as well. In the other languages we measure, dialog-prompting methods appear to control toxic language harms more effectively.}

\subsection{\rev{Interview Guide}}
\label{interview}

\noindent \textbf{Themes to focus:}
\begin{enumerate}
\item General perceptions
 \item Identifying assumptions
    \item Handling/Using assumptions
\end{enumerate}

\smallskip
\noindent \textbf{General perceptions:}
\begin{enumerate}
    \item What is your role, and what do you work on specifically for AI?
    \item Throughout the lifecycle, what are the things that are given, taken for granted, unstated reasons?
    \begin{enumerate}
        \item Why do you think they are given?
        \item How do these "givens" influence how you do your work?
        \item Which of these are easy to identify and handle, and which are not?
    \end{enumerate}
    \item Which stage of the ML lifecycle do you focus more or less on while identifying and handling these “taken-for-granted”? Why?
    \item What is an assumption, according to you? Why? 
    \item In your work, what do you characterize as limitations?
    \begin{enumerate}
        \item Why are these limitations? Examples?
        \item Contextual: What assumptions do you make that inform these limitations?
        \item How do these limitations influence your work?
    \end{enumerate} 
    \item How do you limit or cap the number of assumptions you can identify and handle?
\end{enumerate}

\smallskip
\noindent \textbf{Identifying assumptions:}
\begin{enumerate}
    \item What tools or methods do you use to KNOW that you are assuming something?
    \item How separate is the assumption identification and handling process from your typical workflow?
    \begin{enumerate}
        \item How much effort did you put into finding and handling the assumptions?
        \item Have you used any Responsible AI-related artifacts (documentation, toolkits, etc.) to elicit or discuss assumptions? 
    \end{enumerate}
    \item How difficult is it to examine your or others’ assumptions and deal with them?
    \item How would you identify and discuss someone else’s assumption that is relevant to your work?
    \begin{enumerate}
        \item If they need to make that assumption, what action of theirs will convince you? How would you evaluate their response?
    \end{enumerate}
    \item Have you observed anything where assumptions are craftily embedded somehow? Can you give some examples?
    \item Have you ever wondered how you missed an assumption after you have already missed it?
\end{enumerate}

\smallskip
\noindent \textbf{Handling/Using assumptions:}
\begin{enumerate}
    \item After you identify an assumption, what do you do?
    \begin{enumerate}
        \item Do you inspect how it influences your argument or analysis? What do you do?
    \end{enumerate}
    \item How often do you explicitly use an assumption in your work? Can you give some examples? Why do you think they are necessary?
    \item When you explicitly formulate (come up with or use) an assumption, how do you check if you have installed the assumption accurately? Can you give any examples?
    \item When you need an assumption for your argument, how do you justify or evaluate it? Do you do this explicitly or implicitly?
    \begin{enumerate}
        \item If your assumption does not have a clear or universally acceptable justification, then what do you do?
    \end{enumerate}
    \item How do you explain your assumptions to someone who needs to understand your work? 
    \begin{enumerate}
        \item How do you imagine this process should go, and how does it actually go? 
    \end{enumerate}
    \item When you make a decision, do you think about the assumptions that accompany it? If so, how?
    \begin{enumerate}
        \item Have you changed your decisions after analyzing an assumption? What does the process look like?
    \end{enumerate}
    \item Do you make any distinctions between the assumptions? If so, what are they, and why do you do that? If not, why do you treat all assumptions homogenously?
\end{enumerate}

\smallskip
\noindent \textbf{Case studies:}
\begin{enumerate}
    \item Present the case and explain the situation
    \item What do they think about the text? (open-ended)
    \item How will they articulate their assumptions? (reflects how they identify and use)
    \item How should others have articulated the identified assumptions?
    \item What more is required to better articulate? List form vs. any other forms? Will detailed annotation help?
\end{enumerate}